# Handwriting Biometrics: Applications and Future Trends in e-Security and e-Health

**Author Identity Information**

Marcos Faundez-Zanuy is Full Professor, School of Engineering Tecnocampus, Pompeu Fabra University, Spain.
Email: faundez@tecnocampus.cat

Julian Fierrez is Associate Professor, School of Engineering, Universidad Autonoma de Madrid, Spain.
Email: julian.fierrez@uam.es

Miguel A. Ferrer is Full Professor, Instituto Universitario para el Desarrollo Tecnológico y la Innovación en Comunicaciones, Universidad de Las Palmas de Gran Canaria, Spain.
Email: mferrer@idetic.eu

Moises Diaz is Associate Professor. Universidad del Atlántico Medio, Las Palmas de Gran Canaria, Spain. Email: moises.diaz@atlanticomedio.es

R. Tolosana is Postdoc Researcher, School of Engineering, Universidad Autonoma de Madrid, Spain.
Email: ruben.tolosana@uam.es

Réjean Plamondon is Full Professor, Polytechnique Montréal, Montréal, Canada.
Email: rejean.plamondon@polymtl.ca

# Handwriting Biometrics: Applications and Future Trends in e-Security and e-Health

Marcos Faundez-Zanuy[1]
Tecnocampus, Pompeu Fabra University, SPAIN
faundez@tecnocampus.cat

Julian Fierrez
Universidad Autonoma de Madrid, SPAIN
julian.fierrez@uam.es

Miguel A. Ferrer
Universidad de Las Palmas de Gran Canaria, SPAIN
miguelangel.ferrer@ulpgc.es

Moises Diaz
Universidad del Atlántico Medio, SPAIN
moises.diaz@atlanticomedio.es

Ruben Tolosana
Universidad Autonoma de Madrid, SPAIN
ruben.tolosana@uam.es

Réjean Plamondon
Polytechnique Montréal, Montréal Canada
rejean.plamondon@polymtl.ca

**Abstract**
**Background-** This paper summarizes the state-of-the-art and applications based on online handwritting signals with special emphasis on e-security and e-health fields. **Methods-** In particular, we focus on the main achievements and challenges that should be addressed by the scientific community, providing a guide document for future research. **Conclusions-** Among all the points discussed in this article, we remark the importance of considering security, health, and metadata from a joint perspective. This is especially critical due to the double use possibilities of these behavioral signals.

**Introduction**
On-line handwriting biometrics systems belong to behavioral biometrics as they are based on actions performed by a specific subject [87]. This is complementary to morphological biometrics, which are based on direct measurements of physical traits of the human body [110][87]. From a human behavior and health condition perspective, on-line handwriting biometrics are more appealing and informative than other popular biometrics traits such as fingerprints or iris [14][72]. Although health applications based on on-line handwriting have not been explored in-depth yet, there is a considerable set of possibilities that will probably be developed in the near future, such as diagnosis/monitoring of depression, neurological diseases, drug abuse, etc. [89]. It can be noted that nowadays, most of the published research in biometric signal processing is based on image and speech. This might be linked to the fact that these signals are simpler and cheaper to acquire compared with others such as on-line handwriting where specific digitizing tablets are needed. Fortunately, on-line handwriting signals are more present in our society than a few years ago due to the increasing popularity of tactile devices and their corresponding cost reduction [97]. Thus, the price of the acquisition device is no longer a drawback nowadays. Also, most Commercial Off-The-Shelf (COTS) smartphone devices incorporate handwriting capabilities. As a result, we forecast a considerable growth in applications in this field. Indeed, in a near future these pen-based hand-held devices (Tablets, Phablets, Cellphones…) will help to protect people's sensitive data protection, for example incorporating automatic signature verification and writer identification. Moreover, the same devices could be used to monitor user fine motor control to detect stress, aging and health problems [89]. In this dual context, privacy issues will have to be seriously investigated. This will be especially important when the double use is possible. Ideally, samples acquired for security purposes (e.g., grant access to a facility or authorize a payment) should not be able to reveal health information of the user. Conversely, samples acquired for health monitoring should be anonymized and not convey user identity information [64]. Figure 1 shows an overview of the main application scenarios of biometric handwriting in both security and health fields. It is worth pointing out that this is not a biometric specific problem. For instance, in [60] similar problems are discussed in relation to loyalty cards provided by supermarkets. The advantage for the customer is that he gets discounts and points that can be exchanged for prizes. The advantage for the company is that it can rationalize the products' distribution along the supermarket, and take better care of their prices' politics, stock, etc. However, consumers are not aware about the risk of permitting that the supermarket stores this information. A secondary/double use could be other companies willing to pay for information about which kind of products is purchased by a given person (tobacco, alcohol, drugs, etc.), which kinds of videos are rented, how much money spends each month, etc. Doubtless, consumption habits are more personal and private than some biometric traits, and special care must be taken for this information, although the offers of the companies seem quite tentative.

---

[1] Corresponding author

The challenges and opportunities of handwriting biometrics for e-security and e-health outlined before are particularly well suited to be advanced in the framework of Cognitive Computing. Combined work considering at the same time both disciplines (Biometrics and Cognitive Computing) can be seen in a few selected works in the past [31][61][74][6], but still many synergies between them are to be exploited in future research and development.

According to the Cognitive Computing Consortium, cognitive computing systems must have five key attributes, as listed below [22][38]:
- Adaptive: Cognitive systems must be flexible enough to learn as information changes and as goals evolve. The systems must be able to make adjustments as the data and environment change, such as user stress.
- Interactive: Human-computer interaction (HCI) is a critical component in cognitive systems. Users must be able to interact with cognitive machines and define their needs as those needs change, maybe due to aging or temporary health state.
- Iterative and stateful: Cognitive computing technologies can also identify problems by asking questions or pulling in additional data, if a stated problem is vague or incomplete.
- Contextual: Understanding context is critical in thought processes, and so cognitive systems must also understand, identify and mine contextual data. They may draw on multiple sources of information present in a given handwriting task.

A handwriting biometric system should be aware and take advantage of the rich amount of information that can be extracted from a very simple handwriting task, as described in the next sections. This is because handwriting is a cognitive task in which synchronized neuromotor orders are fired from the cortex to carry out the planned action [5]. Signal processing and pattern recognition techniques applied to analyze this procedure can reveal very useful information about the writer, including its identity and health state.

The present paper summarizes the current research in handwriting biometrics with applications to e-security and e-health under the umbrella of Cognitive Computing principles: adaptiveness, interactivity, iterativity, and context. We focus our discussion on the accuracy, performance, utility, application factors, and challenges of handwriting analysis in such application scenarios, also foreseeing trends and potential research lines for the future. More technical surveys of the state-of-the-art of handwriting analysis for security and health can be found elsewhere [59][14][37][89].

**The Biometric Data Behind Handwriting**

The acquisition devices considered in on-line handwriting allow to capture various properties of the moving pen (or fingertip [102]) during the whole writing process in real-time. These time-stamped sequences of points typically give the digital representation of the signal. For instance, digitizing tablets typically acquire the following information:
- position of pen tip in X-axis,
- position of pen tip in Y-axis,
- on-surface/in-air pen position information,
- pressure applied by the pen tip,
- azimuth angle of the pen with respect to the tablet's surface,
- altitude angle (a.k.a. tilt) of the pen with respect to the tablet's surface,
- timestamp.

An example of this information is given in Figure 2 for both healthy and unhealthy handwriting.

From this set of dynamic raw data, further information can be inferred or derived, which is usually more suitable for certain applications (e.g. handwriting velocity, duration, width, height) [71]. Some digitizing devices such as Wacom Intuos tablet, Samsung Galaxy Note, etc. are able to track the pen tip movement even when it is not touching the surface (see Figure 3). Thus, it is possible to record the X and Y coordinates of in-air movements when pressure is null. Unfortunately, this is only possible when the distance between the tip of the pen and the surface is less or equal to approximately 1 cm; otherwise the tracking is lost. Nevertheless, the in-air time spent is still known because the acquisition device provides a timestamp for each sample [94]. In-air information cannot be obtained with "classic" off-line methods (scanning the ink-pen handwriting once the task is finished) and the timing information is also lost or very difficult to recover [34][86].

An accurate complete recording of in-air information requires other 3D sensors. Possibly, the more appropriate for handwriting are the Motion Capture (MOCAP) sensors. These sensors are based on either video or inertial measurement units (IMUs), i.e. accelerometer, gyroscope and magnetometer. The first ones, e.g. Vicon system (www.vicon.com), use the image from multiple cameras to calculate the 3D position of the target, which could be the tip of the pencil or a fingertip, where a marker is attached. The second ones are based on IMUs, such as neuronmocap (https://neuronmocap.com), which require a glove in which sensors are plugged in to track the 3D trajectory. In both cases, it is possible to sample the position of the marker or IMUs around 100 times per second (i.e. 100 Hz). The major problem of the first ones is the occlusion whilst the downside of the second is the offset drift of the accelerometers. Besides, the first one must be used in a fixed space while the second is portable. Anyway, both procedures are reliable systems for 3D tracking which, although not massively used in nowadays handwriting biometric applications, is a promising research line to be explored. We have everything in hand to design 2D and 3D systems from the acquisition hardware point of view. While 2D models and algorithms are a reality, the potential of 3D systems is yet to be explored.

**Security Applications**

According to [35], verification and recognition are two classical studies carried out with handwriting specimens. In the context of biometric security, applications based on handwriting tasks are mainly based on signatures. Several international competitions summarize the state of the art achieved by dozens of teams, such as BSEC [77], SVC [28] and SigWIComp [65]. Although less known, there are also some studies (e.g. [11][109][15]) where biometric recognition is based on handwritten text, either text-dependent or independent.

Several authors have demonstrated the individuality of handwriting. It has been assessed in the off-line case for example in [88]. These researchers collected a database of 1500 writers selected to be representative of the US population and conducted experiments on identification and verification. Regarding writer identification, they reached an accuracy of about 83 % at the word-level (88 % at the paragraph level and 98 % at the document-level). These results allowed the authors to conclude that the individuality hypothesis, with respect to the target population, was true with a 95 % confidence-level. Another study [11] complemented the previous work of [109]. They analyzed the individuality of four handwritten words (*been*, *Cohen*, *Medical* and *referred*) taken from 1027 US individuals who wrote each word three times. The combination of the four words yielded an identification accuracy of about 83 % and a verification accuracy of about 91 %.

With regard to the on-line case, some authors have addressed the issue of the individuality of single words and short sentences. [15] showed that single words (the German words *auch*, *oder*, *bitte* and *weit*) and the short sentence *Guten Morgen* exhibit both considerable reproducibility and uniqueness (i.e. equal items written by the same person match well while equal items written by different people do not match so well). They used a small database consisting of 15 writers that produced, in a single session, ten repetitions of each item captured by a prototype of a digitizing pen. In [39] the English words *February*, *January*, *November*, *October* and *September* were used (25 repetitions of each word donated by 45 writers). The identification rate reached 95 %. In [29] a writer identification rate of 92.38 % and a minimum detection cost function of 0.046 (4.6 %) was achieved with 370 users using just one word written in capital letters. Results were improved up to 96.46 % and 0.033 (3.3 %) when combining two words. In [75], signatures, initials and keywords were compared.

More recently, in [102] writer verification rates of 96.2 % were achieved for handwritten passwords consisting of 7 digits written with the fingertip over the touchscreen of COTS smartphones and 94.1 % for 4-digit passwords. In a subsequent study, the same authors were able to obtain similar authentication accuracy for more challenging and realistic acquisition conditions (in the wild including cross-device comparisons over a diverse pool of COST smartphones consisting of 94 different models) by considering characters and symbols instead of only numerical digits to construct the passwords [99].

In a separate line of research, some authors have also explored handwriting information from drawings and other graphical information generated with stylus- or finger-based inputs for authentication [70], usually resulting in worse performance as compared to natural handwriting or signature data [69]. Anyway, the results obtained based on graphical inputs may be acceptable for some low-security applications, or helpful as a second authentication factor (e.g., in smartphone login applications [2]).

It is also worth pointing out some forensic challenges approached mainly in off-line handwriting. The two main tasks addressed by forensics are forgery detection and disguising. Forgeries refer to a writer imitating the handwriting of other writers while disguising means a writer trying to conceal his usual writing habits and to make the new writing as different as possible to his own writing [96]. An open question is, knowing that a handwriting specimen is a forgery, how to detect who has forged the handwriting. Some attempts have been made [56] to track this problem but with poor results. Regarding disguising, this problem has also attracted much attention in the forensic area. This problem arises when the author changes his own handwriting. The purpose could be for the later denial of the own handwriting, in which case it is also termed as "auto-simulation". Note that both genuine and disguised texts are written by the author of the specimen but with different intentions. A writer produces a genuine text that allows author identification. Whereas, a disguised handwritten is written to make it look like a forgery for a possible later denial. These cases have been mainly studied in handwriting signatures [68][52]. The difficulty of forensic-based challenges can be faced by on-line systems. Since they have the advantage of having the timing information, it is expected better results in a near future.

**Health Applications**

Although most of the many existing studies related to handwriting and handwritten signatures have been based on on-surface movements (see for example [87][59]), some studies have pointed out the importance of in-air movements as well. In [30] the authors performed an entropy analysis of handwriting samples acquired from a group of 100 people (for more information refer to BiosecurID database [41]) and observed that both types of movements contain approximately the same amount of information. Moreover, based on the values of mutual information, these movements appear to be notably non-redundant. This property has been advantageously used in several fields of science. For instance, the authors in [79][80] proved that in-air movement increases the accuracy of Parkinsonic dysgraphia identification. Specifically, when classifying the Parkinsonic dysgraphia by Support Vector Machines (SVM) in combination with the in-air features, they reached 84% accuracy which is 6% better in comparison to classification based on on-surface features only. When combining both feature sets, they observed 86% classification accuracy. Similar accuracy rates were obtained in a different study by [83], which analyzed various handwriting tasks to discriminate between Parkinson Disease (PD) vs. Elder Healthy Controls. Better accuracy rates around 96 % were also obtained in [83] when comparing PD vs. Young Healthy Controls. Effective accuracy rates are also found when Convolutional Neural Networks (CNN) are used in image-based sample representation of in-air and on-surface traces in [57][73].

The in-air movement also supports the diagnosis of Alzheimer's disease (AD) [50]. These researchers observed that patients with AD spend 7-times longer in-air when compared to a control group. In the case of on-surface movement, it is only 3-times longer. Similarly, [107] found out that the in-air duration can be a good measure for performance analysis of children with a high-functioning autism spectrum disorder. The in-air movement has also been used for the identification and the quantitative analysis of developmental dysgraphia in child population [46][108][106]. In [46], it was proved that kinematic features derived from this kind of movement (especially jerk, which is the rate at which the acceleration of a pen changes with time) provide good discrimination power between children with dysgraphia and the control group.

On the other hand, there is a successful research line in handwriting analysis for health applications based on kinematic models. There exist many theories that have tried to model the speed profile of human movement in general and handwriting in particular [88]. Among the models which provide analytical representations, the Kinematic Theory of rapid human movements [91][92][93] and its Delta and Sigma–Lognormal models have been used to explain most of the basic phenomena reported in classical studies on human motor control [85] and to study several factors involved in the fine motricity [9].

As can be seen in Table 1, the Sigma-Lognormal model represents a breakthrough due to its feasibility and reliability to describe a wide range of human movements. One of the main advantages of this model is that it considers physical body features such as the neuromuscular system responsible for the production of human movements and thus reflects some personal characteristics difficult to impersonate. To work out the Sigma-Lognormal parameters, the Robust Xzero algorithm implemented by the ScriptStudio application was proposed in 2007 [18] and iDeLog was proposed in 2020 [55]. Since its first version, it was quickly widespread, and several improvements have been published, for instance [27]. From a fundamental perspective, the powerfulness of this methodology relies on the lognormality principle [90] which states that the lognormal velocity profiles observed in handwriting and signature reflect the behavior of individuals who are in perfect control of their movements. As a corollary, motor control learning in young children can be interpreted as a migration toward lognormality while, as aging and health issues intensify, a progressive departure from lognormality is occurring. For the greater part of their lives, healthy human adults take advantage of their lognormality to control their movements efficiently [94].

For a recent review of handwriting analysis to support neurodegenerative diseases diagnosis see [14][37][89].

**Metadata Applications**

Behavioral biometrics, in addition to security and health applications, can provide additional information, known as metadata. Sometimes also referred to as Soft Biometrics [82], they can be based on system hardware specifics (technical metadata) or on personal attributes (non-technical metadata) [112][20]. System-related metadata represent the physical characteristics of biometric sensors and are essential for ensuring the minimum acceptable quality of the raw biometric signals. Previous work in personal related metadata has shown that it is possible to estimate some metadata like script language, dialects, origin, gender, and age by statistically analyzing human handwriting.

As examples of gender recognition based on handwriting, we can mention the following works. In [31] using only four repetitions of a single uppercase word, the average rate of well-classified writers is 68 %; with sixteen words, the rate rises to an average of 72.6 %. Statistical analysis reveals that the rates mentioned above are highly significant. In order to explore the classification potential of the in-air strokes, these were also considered. In that case, results were not conclusive, with an average of 74 % well-classified writers when information from in-air strokes was combined with information from on-surface ones. This rate is slightly better than the one achieved by calligraphic experts. However, we should keep in mind that this is a two-class problem and even by pure chance we would get 50 % accuracy.

A system that classifies handwriting into demographic categories using measurements such as pen pressure, writing movement, stroke formation, and word proportion has been proposed in [39]. The authors reported classification accuracies of 77.5 %, 86.6 %, and 74.4 % for gender, age and handedness classification, respectively. In this study, all the writers produced the same letter. Authors in [66] also addressed the classification of gender and handedness in the on-line mode. The authors used a set of 29 features extracted from both on-line information and its off-line representation and applied SVM and Gaussian Mixture Models (GMM) to perform the classification. The authors reported accuracy of 67.06 % for gender classification and 84.66 % for handedness classification. In [67], the researchers separately reported the performance of the off-line mode, the on-line mode, and their combination. The accuracy for the off-line mode was 55.39 %.

Emotional states, such as anxiety, depression, and stress, can be assessed by the Depression Anxiety Stress Scales (DASS) questionnaire. A new database that relates emotional states to handwriting and drawing tasks acquired with a digitizing tablet has been presented in [51]. Experimental results show that the recognition of anxiety and stress was better than depression recognition. This database includes samples of 129 participants whose emotional states were assessed by the DASS questionnaire and is freely distributed for those interested in research in this line.

**Applications Combining Security and Health: Issues and Challenges**

Security and health applications have been discussed in previous sections as fields isolated from each other. They are indeed studied separately for most of the scientific community. However, in some cases, both should be fully considered jointly as they cannot be completely isolated one from the other (see Figure 4). This is particularly true in the context of the Personnal Digital Bodyguard vision [89] which aims at supplementing people's sensitive data protection with signature and writer verification.

Most biometric security applications only try to determine the identity of a subject or to verify if he/she is the person who claims to be. However, in the context of the acquisition of biometric signals with different devices and data quality [33][96], it may be beneficial to gather some additional information for the system to be context-aware [74], as envisioned by the Cognitive Computation initiative. In the following, we summarize three of such possible scenarios:

a) Is the subject stressed? It is not the same problem to confirm the identity and open a door if, for example, the subject's heart is beating at 70 beats per minute (bpm) or if it is beating at 120 bpm. If the heart is much more accelerated than normal, some suspicious activity can be happening (e.g., the user is being coerced). To solve this, some biometric systems have a mechanism to notify security guards or the police about the coerced situation without letting the threatening person notice it. To do that, the user may have enrolled at least two different biometric templates. Both will open the door, but one of them will activate a silent alarm. This concept is known as duress detection [16]. This knowledge can also be obtained just considering how the subject interacted with the sensor in previous days. Similarly, the user can enroll in a couple of different signatures, one for duress recognition and the other one, for a normal operation. Again, it would be possible for a third party to be familiar with the genuine signature that does not activate any silent alarm and to force the user to use that signature.

   A robust biometric security system should be able to detect the stress situation based on characteristics that cannot be easily controlled by the user. Detection of user stress from signature or handwriting is a challenging research topic that can indeed improve security systems.

b) Is the subject suffering from any disease that makes him unable to understand the real implications of his acts? In [61], we presented the case of a woman affected by Alzheimer's disease. In that case, several women made an elderly woman sign her name on blank sheets of paper, alleging that they needed her signature to get some of her medicines. When the elder woman died, the women took advantage of the signed sheets in order to write a rental agreement. The declared date of this agreement was 1985, but several documents signed in 1986 showed better control of calligraphic movements. In fact, the hesitantly written signature document signed in 1985, was closer in appearance to the blank sheets signed when the elderly woman had dementia than to the 1986 document. Thus, it was demonstrated that in fact, the rental document was not signed in 1985 but later. While this is an off-line example from the 80's of the past century, we can forecast in a near future situations like this one for the on-line case.

   Another possibility is to be affected by depression. Drawings for analyzing depressive disorders in older people were used in [34].

   These two examples indicate that even if in the context of biometric signature verification one can conclude that the signature is genuine, this may not be enough. One may in addition take into account aspects such as the health state of the subject. Considering both aspects (identity and health) it can be concluded that, from a legal point of view, a signature is not valid because the subject may not have been in a sound state of mind. In such situations, the biometric authentication of the individual does not solve the problem and some additional considerations should be taken into account. This is not just related to health. Another similar situation where a genuine biometric sample is used fraudulently is a replay attack. In a replay attack, the biometric signal is usually genuine, but it was acquired/recorded in the past and presented again and should be considered as a fake attempt [105].

c) Is the individual temporarily affected by drug substances? Changes in handwriting due to alcohol were reported in [3][42]. The effects of caffeine on handwriting were detected in [78]. Similar experiments regarding the effects of marijuana and alcohol were performed in [70].

One of the main concerns of biometrics applied to security is privacy [60]. Technological advances allows to store, gather, and compare a wide range of information on people. Using identifiers such as name, address, passport or social security number, institutions can search databases for individuals' information. This information can be related to salary, employment, sexual preferences, religion, consumption habits, medical history, etc. This information can be collected with the consent of the user, but in some cases, it could also be extracted from biometric samples without the knowledge of the user. Thus, the user could be completely unaware that some additional and private information can be extracted from his biometric samples [7]. Therefore, there is a potential risk.

Let us think, for instance, in sharing medical information. Obviously, in the case of an emergency, this sharing among hospitals would be beneficial. On the contrary, if this information is transferred to a personal insurance company or a prospective employer, the insurance or the job application can be denied. The situation is especially dramatic when biometric data collection is intended for security applications to grant access to a facility or classified information, but a third party tries to infer the health condition of the subject. For instance, in the case of eye biometrics [32], an expert can determine that a patient has diabetes, arteriosclerosis, hypertension, etc.

For any biometric identifier, there is a portion of the population for which it is possible to extract relevant information about their health, with similar implications to the ones described in the previous paragraph. This is not a specific problem of handwriting signals. Some other biometric signals exhibit the same potential problems. For example, speech disorders, hair or skin color problems, etc. An important question is what is exactly disclosed when biometric scanning is used. In some cases, additional information not related to identification might be obtained. One possible scenario could be a company where an attendance system requires workers to sign each day. The main purpose of this task could be to check if the worker is at his workplace during the working days. However, once the handwriting is provided, the company could decide to analyze the signature to detect some pathologies or drug abuse and to dismiss those workers who do not show good health. And last but not least, once we provide our biometric samples they can stay in a database for dozens of years and due to technological advances, they can be used in a simple way in the future to extract additional information that was not intended during acquisition. For this reason, we should think about technical solutions to preserve privacy and legal regulations to avoid such foreseeable issues.

Sometimes the situation is just the opposite. With the growth of interest in the fields of e-health and telemedicine, scientists started to develop automatic handwriting analysis systems that can be used for disease diagnosis [36], rating, or monitoring [94]. Basically, a database of healthy control samples from healthy individuals and pathological samples is acquired, disseminated and used for research purposes. Its registers are usually anonymized removing (destroying) any link to donor names and/or passports or national identity codes. However, this anonymization process could be useless if the acquired samples permit a biometric identification based on handwriting or drawings.

This might not be feasible at present, but it might be possible in the near future with the improvement of the processing algorithms.

A related topic is de-identification for privacy protection in multimedia content. De-identification in multimedia content can be defined as the process of concealing the identities of individuals captured in a given set of data (images, video, audio, text), to protect their privacy. This will provide an effective means for supporting directives and laws related to personal data protection, e.g., the EU's General Data Protection Regulation (GDPR).

**Future Trends**

Security and health applications of handwriting analysis cannot be considered as separate fields anymore. While significant scientific developments exist in the field of biometric recognition using the handwritten signature for security [72][59], few efforts have been made in general handwriting [102] and drawing [69] for security. Few efforts have also been deployed in health applications [36][83][14][37] as well as in possibilities of combining security and health.

As a summary of the research discussed in the present paper, on the one hand, the state-of-the-art in applications based on online handwritten tasks can be considered:

a) mature in security applications based on signature [72][59],
b) incipient in security applications based on the handwritten text [102][99],
c) incipient in security applications based on drawing tasks [70][69],
d) incipient in health applications based on handwriting text and drawing tasks [21][83][14],
e) incipient in health applications based on signature [83][59][14][36].

On the other hand, it is difficult to find published research in joint applications where security is studied in combination with one of the following health-related aspects:

f) is the user to authenticate under stress?
g) is he suffering any disease that makes him unable to understand the real implication on his acts?
h) is he temporarily affected by drug or substance abuse?

Similarly, there is a lack of research in applications where health is the main goal and the following issue is addressed at the same time:

i) In order to keep the user's privacy [64], the handwritten signal is transformed [7] in order to remove his identity while preserving the diagnosis potential of the sample.

For the future, we see four main areas of research in online handwriting analysis for security and health, at shown in Table 2:

- Processing, analysis, and recognition of handwriting signals; with special emphasis in motor models based on neuroscience [47] and methods based on deep learning [98][97] able to make the most of large-scale datasets [95].
- Support for early diagnosis [36] of pathologies using handwriting, especially neurodegenerative diseases [83][37][14].
- Continuously monitoring the evolution of neurodegenerative diseases, rehabilitation, and healthy aging [98][99]; using handwriting and other touch interaction signals [40].
- Removing undesired information from handwriting representations [7]: removal of health information for security applications and removing of identifying information in tasks acquired for health analysis purposes [64].

These four areas of future research can be further broken down as follows:

1. Development and scientific progress in biometric security applications based on handwritten text (capital letters, cursive letters, etc.) [102][99], where the goal is not the classical Optical Character Recognition (OCR). The goal is to identify the author of the text in two different modes: text-dependent and text-independent. Additional functionality is checking if the author of the text is forced to write a specific text or if he is free to write whatever he wants.

2. While signature-based recognition systems are quite mature, the complementary information of the combination between signature and text provided by the same author as well as the possibility of crossed recognition has not been deeply analyzed yet.

3. Development of new algorithms able to identify the author of a drawing in a similar way to what is done with signatures or handwriting. In the same way that a specific simple pattern can be used to unlock the smartphone [2], a specific drawing (invented or copied) could hold the same functionality [69]. Pattern unlocks consist of a grid of dots on a device's lock screen which users connect uniquely to gain access to the phone. However, the possibilities are much reduced when compared to handwritten drawings.

4. Development and scientific progress in biometric health applications based on handwritten text and drawings [83]. This is a promising research line to detect healthy aging, evaluate the effect of prescribed drugs on a specific disease (such as apomorphine on Parkinson's Disease), etc.

5. Analysis of the potential use of handwritten signatures for diagnosing diseases [36][83]. Although the signature tends to be a mechanical movement with almost no cognitive effort to perform it, it has potential use in health applications.

6. Stress detection given handwritten tasks has potential implications on security applications, e.g., coercion detection [96].

7. Drug substance abuse can have legal implications for professional activities. It would be cheap and non-invasive to conduct preliminary tests based on handwriting tasks.

8. Development of algorithms to anonymize handwritten samples in order to hide the identity of the user while preserving the diagnostic capability of the samples [7].

9. Development of algorithms to anonymize handwritten samples in order to hide the health information of the user while preserving the identification capability of the samples [64].

In order to advance these research lines, adequate and realistic handwritten data is needed for experimentation. That data should necessarily involve several acquisition sessions, approved by an ethical committee, in order to cope with the temporal variability of the users as well as different tasks:

- Signature;
- cursive letters (free text and predefined text);
- capital letters and digits (free text and predefined text); and
- drawings (simple strokes, Archimedes spiral, clock drawing tests, pentagon tests, house copying tests, etc.)

Fortunately, some databases are already available for research in several scenarios:

- Biometric security using signature [59][48][95][98] and text [41] [99];
- Parkinson's disease and essential tremor [14][83][21];
- Alzheimer's disease and mild cognitive impairment [93]; and

- Dysgraphia [35].

**Conclusion**

In this paper we have pointed out some future trends and challenges in biometric research on signature and handwriting. Special emphasis is given to the fact that, contrary to other biometric traits, handwriting signals are of interest in both, e-security and e-health. Some challenges are identified and should attract the interest of research community to live in a secure society.

We consider that this paper could help future researchers on online handwritting analysis to identify the main research topics to be addressed in the next years, as well as to attract new researchers from other fields.

**Acknowledgements**
This study was funded by Spanish grants MINECO/FEDER TEC2016-77791-C4 and RTI2018-101248-B-I00, Bio-Guard (Ayudas Fundacion BBVA a Equipos de Investigacion Cientifica 2017), and Cecabank. Réjean Plamondon has been supported by NSERC Grant RGPIN-2015-06409. Ruben Tolosana enjoys a postdoc position funded by Comunidad de Madrid (PEJD-2019-POST/TIC-16031).

**Compliance with Ethical Standards**
The authors declare that they have no conflict of interest.
All procedures performed in studies involving human participants were in accordance with the ethical standards of the institutional and/or national research committee and with the 1964 Helsinki declaration and its later amendments or comparable ethical standards. For this type of study formal consent is not required.
This chapter does not contain any studies with animals performed by any of the authors.
Informed consent was obtained from all individual participants included in the study.

# FIGURES AND TABLES

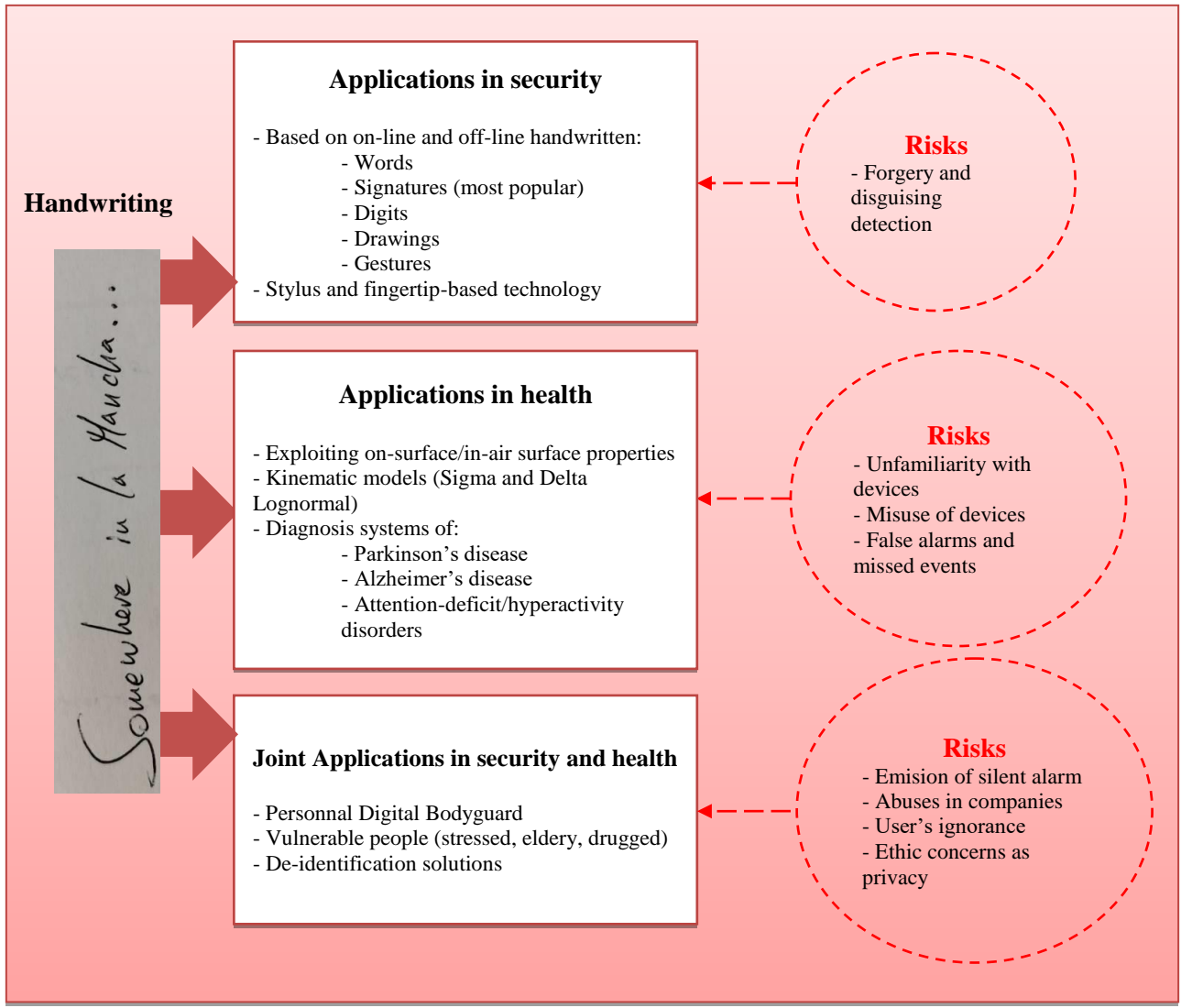

Figure 1: Exploiting on-surface/in-air propierties for biometric handwriting.

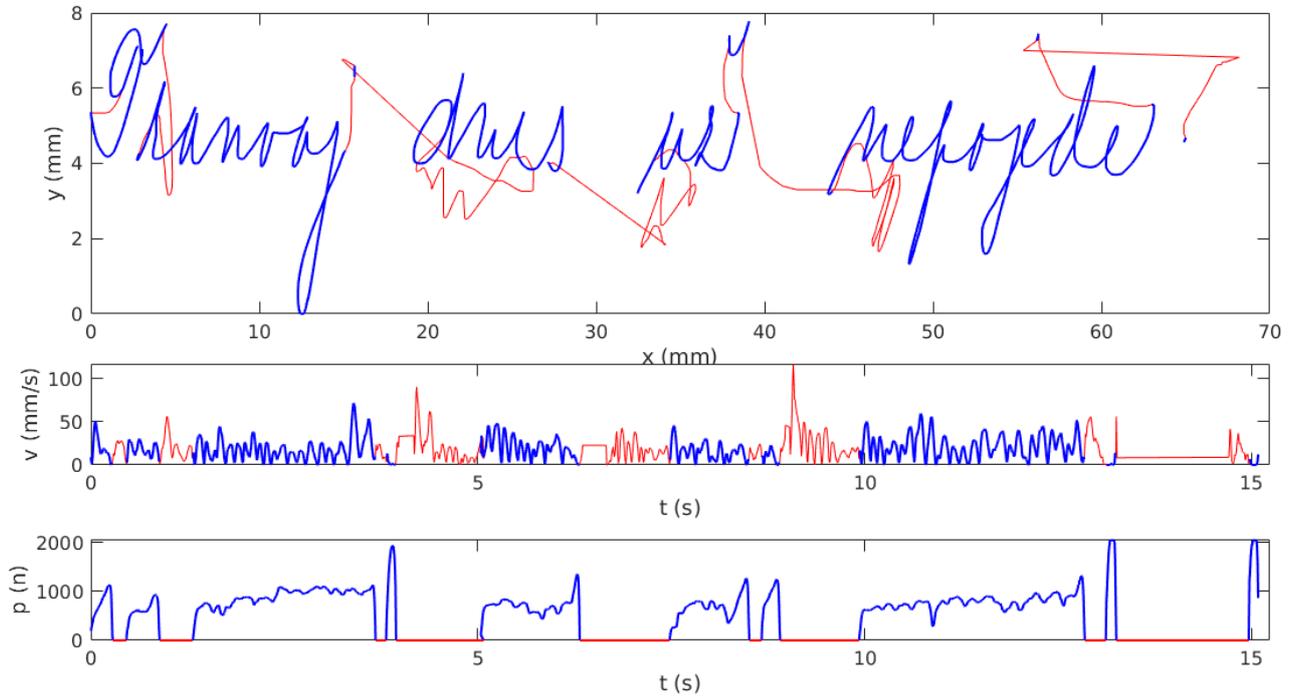
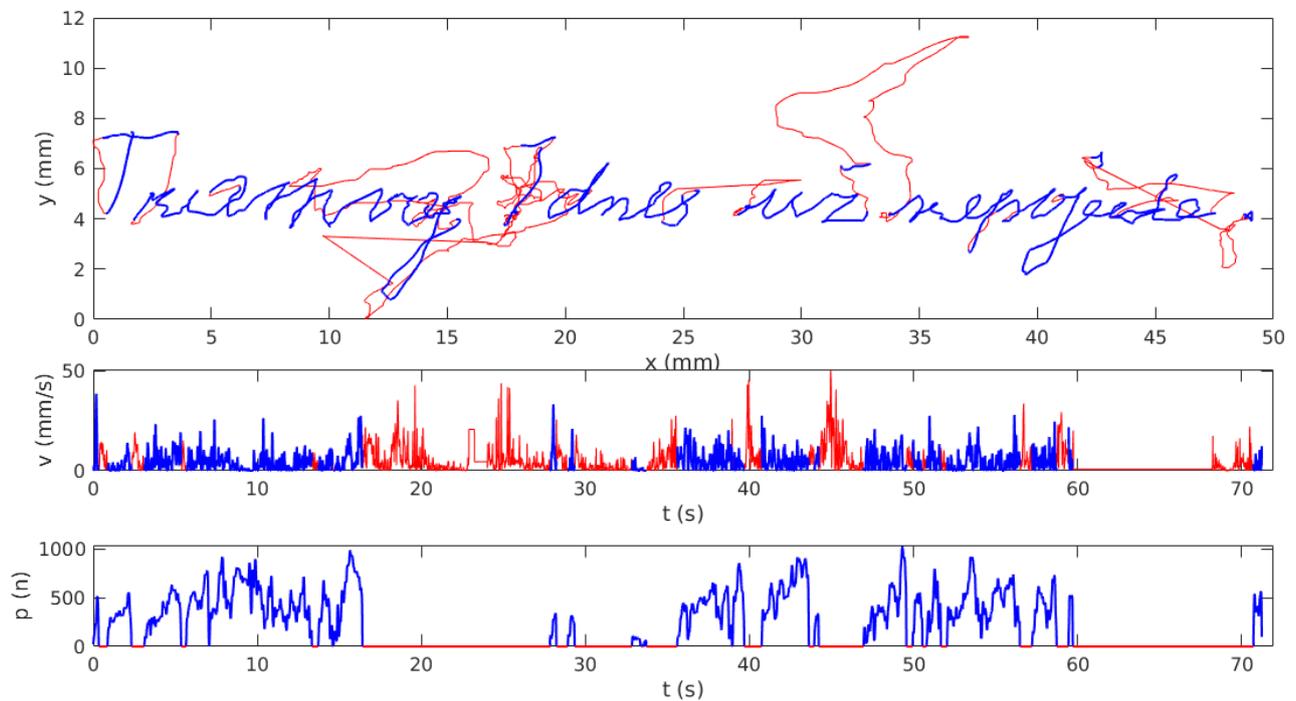

*Figure 2: On-line handwriting from healthy and unhealthy subjects: in-air handwriting in solid red line and on-surface handwriting in solid blue line. (Figure extracted from PaWaH database [79]).*

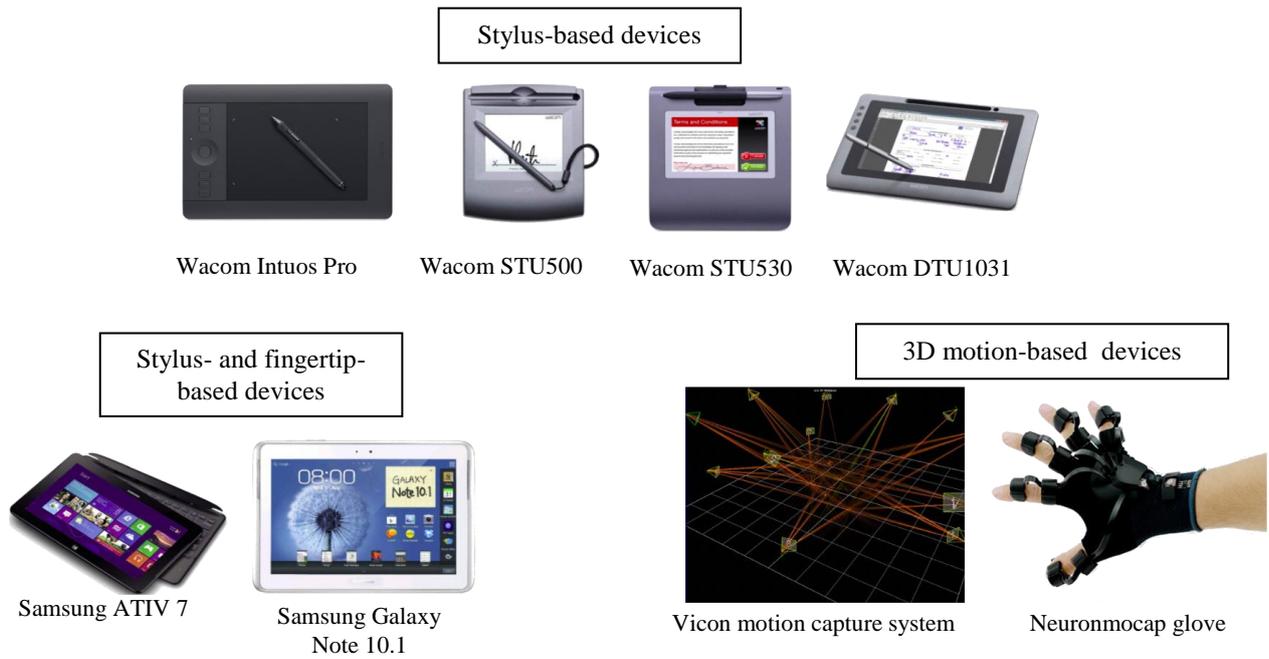

*Figure 3: Devices for capturing handwriting information. Illustration designed using figures extracted from: [97], https://thumb.pccomponentes.com/w-530-530/articles/5/59297/wacom-intuos-pro-small-1.jpg, www.virtualrealityreviewer.com/wp-content/uploads/2014/09/PERCEPTION-NEURON.jpg and https://www.researchgate.net/profile/Ron_Wakkary/publication/221572400/figure/download/fig3/AS:305541982244866@1449858293504/Graphical-view-of-the-Vicon-motion-capture-system-Twelve-cameras-are-used-to-reconstruct.png*

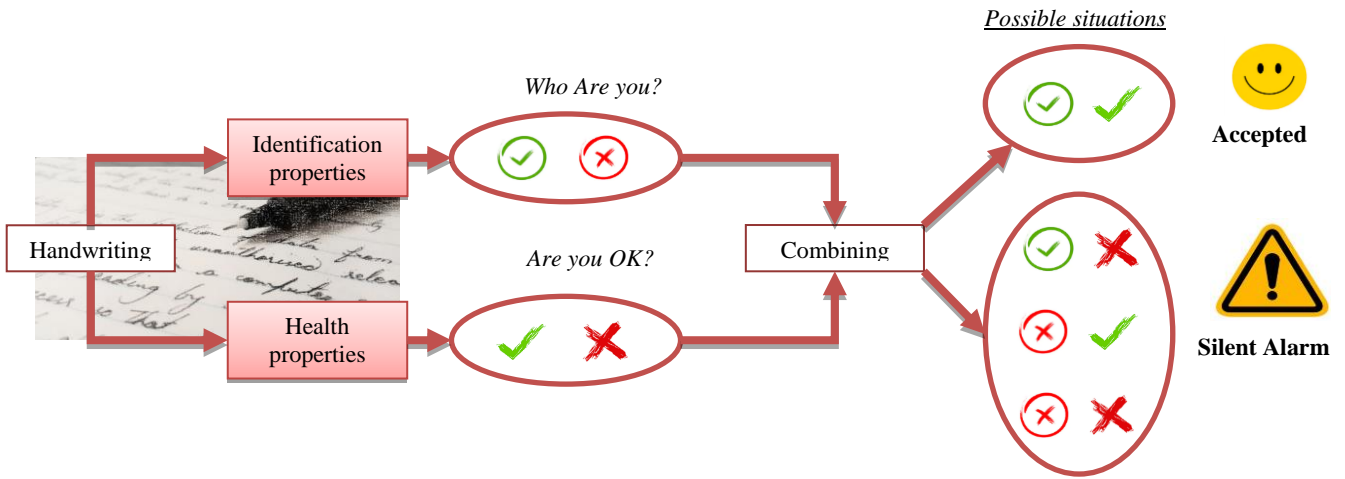

*Figure 4: Towards a combination of identification and health properties of biometric handwriting information.*

# TABLES

*Table 1: Some Kinematic Theory applications related to handwriting biometrics.*

| Health-based applications | Security-based applications | Otherhandwriting-based applications |
|---|---|---|
| Temporal evolution of handwriting skills [13] | Signature verification systems [9][4] | Western [54] and Indian [113] handwriting generation |
| Tools to help children learning handwriting [103] | Training improvements with duplicates data augmentation [58] | Gesture generation [1] |
| Tools for neuromuscular health care [24] | Forgery detection improvements [63] | Captcha generation [19], |
| Motor control disorders and brain strokes-based systems [17] | On-line learning improvements with handwriting generation [76] | Graffities design [23] |
| Parkinson disease-based systems [8] | | Mouse movement analysis [25], |
| Turn cranio-caudal signature characterization [49]. | | Uni and bi manual drawing movements [114] |
| Children with ADHD [81] | | Human machine Interface [53][26] |

*Table 2 : Future developments in the area of handwriting analysis: Research Areas and Specific Problems.*

| Research Areas | Specific Problems/Technologies |
|---|---|
| 1. Fundamental research in:<br>• Neuroscience-based models<br>• Deep learning<br>• Large-scale datasets<br>2. Supporting early diagnosis of pathologies<br>3. Continuous monitoring the evolution of diseases<br>4. Removing either health state or identity information | Writer identification in text-dependent and text-idependent modes |
| | Recognition systems combining signatures + text |
| | Drawing, patterns, and touchscreen interaction for user identification |
| | Novel health-based applications using text, drawings, and touchscreen interaction |
| | Signatures for diagnosis diseases |
| | Detection of stress |
| | Detection of drug substance intakes |
| | Anonymization of identity by keeping diagnostic properties |
| | Removing/obscuring undesired health information by keeping identification properties |
| --------- ***Necessity of handwriting data for working on future areas*** --------- ||